\documentclass[runningheads]{llncs}
\usepackage[T1]{fontenc}
\usepackage{multirow}
\usepackage{xcolor}
\usepackage{booktabs}
\usepackage{graphicx}
\usepackage{hyperref}
\usepackage{orcidlink}

\begin{document}
\title{GreenPipe: Power Modeling for Containerized DNN Inference on Kubernetes Edge Nodes}
\titlerunning{GreenPipe: Power Modeling on Kubernetes Edge Nodes}
\author{Mengxue Wang\inst{1}\orcidlink{0000-0002-8262-9333} \and
Peini Liu\inst{2}\orcidlink{0000-0003-0058-8732} \and
Amir Taherkordi\inst{3}\orcidlink{0000-0003-1672-054X} \and
Jordi Guitart \inst{1,2}\orcidlink{0000-0003-0751-3100}
}
\authorrunning{M. Wang et al.}
%
\institute{Universitat Politècnica de Catalunya, Barcelona, Spain \email{\{mengxue.wang,jordi.guitart\}@upc.edu}  \and
Barcelona Supercomputing Center, Barcelona, Spain \email{\{peini.liu,jordi.guitart\}@bsc.es}
\and
University of Oslo, Oslo, Norway \\
\email{amirhost@iﬁ.uio.no}
}
\maketitle              
\begin{abstract}

Distributed DNN inference is increasingly deployed in containerized edge--cloud environments, where workloads run on-device or are exposed to remote clients over the network.
Accurate online power estimation on resource-constrained ARM nodes without hardware power counters such as RAPL remains a challenge, and CPU-only models fail to capture multi-resource behavior. 
We present \textbf{GreenPipe}, an automated profiling-training-validation pipeline that builds multi-resource regression models from external power meter measurements and attributes power to containers proportionally.
GreenPipe is evaluated on a Raspberry Pi 4 edge node in a K3s edge--cloud testbed, covering DNN inference with three vision models, multiple precisions, thread counts, and both local and serving scenarios.
System-level MAPE is 6.3–9.4\%, improving over CPU-stress and utilization-only baselines by 26.9\% MAPE on average. We jointly report inference latency and energy per inference, exposing performance–energy trade-offs across workload configurations.

\end{abstract}
\section{Introduction}
Distributed DNN inference is increasingly deployed on the edge--cloud continuum, where services run locally on resource-constrained edge nodes or as network-accessible endpoints~\cite{9596610}. These deployments need power visibility for energy-aware orchestration, thermal control, and sustainable operation~\cite{10637271}. However, many ARM edge devices lack hardware power interfaces such as RAPL, while external meters are usually impractical during normal runtime~\cite{10.1016/j.comnet.2024.110371}. Software power models are therefore needed, but CPU-centric models can miss the cache, memory, disk, and network behavior of containerized DNN inference~\cite{10171575}.

Container orchestration adds another challenge: power must be exposed at a granularity useful for workloads. Kubernetes and Docker provide isolation and cgroup-scoped metrics, but node-level meters alone cannot identify the contribution of individual containers~\cite{10.1007/978-3-031-50684-0_20,10.1145/2996890.2996899}. This is particularly relevant for edge--cloud inference, where local execution and network-served inference stress different resource paths and lead to different latency--energy behavior.

This paper presents \textbf{GreenPipe}, a data-driven pipeline for building and deploying per-node power models for containerized DNN inference on Kubernetes ARM edge nodes. GreenPipe collects multi-resource metrics, aligns them with external meter measurements, trains and validates regression models, and deploys the selected model for per-second online estimation with heuristic container-level attribution.
We evaluate GreenPipe on a Raspberry Pi~4 in a two-node K3s testbed, covering three image-classification models, multiple precisions and thread counts, and local versus network-served inference. GreenPipe achieves node-level MAPE of 6.3\%--9.4\% on validation workloads and reduces average MAPE by 26.9\% over CPU-stress and utilization-only baselines. We also characterize latency--energy trade-offs across workload configurations and deployment modes.
Our main contributions are:
\begin{itemize}
    \item a Kubernetes-compatible pipeline for multi-resource profiling, power model training, validation, and online estimation on ARM edge nodes;
    \item DNN-oriented training and validation across local and network-served containerized inference workloads;
    \item runtime node-level estimation with heuristic container attribution, together with latency--energy characterization for deployment decisions.
\end{itemize}

\section{Related Work}

\paragraph{Software-based power modeling on edge platforms.}
Software power models estimate node power from observable resource metrics when direct hardware energy counters are unavailable. Prior works have modeled full-system power using CPU utilization, performance counters, frequency features, and regression-based methods~\cite{10.1145/2996890.2996899,KANSO2023100837,10994218}. For Raspberry Pi devices, Kanso et al.~\cite{KANSO2023100837} train utilization-based models from CPU stress workloads, while Wang et al.~\cite{10994218} study embedded devices using utilization and frequency features. These approaches demonstrate the feasibility of software power estimation, but their training data are often CPU-centric or based on micro-benchmarks, which limits accuracy for DNN inference workloads that also exercise memory, cache, disk, and network resources. GreenPipe follows this software-modeling direction, but trains from meter-labeled multi-resource benchmarks and validates on containerized DNN inference under local and network-served execution.

\paragraph{Container-aware power meters and orchestration frameworks.}
Containerized deployments require power visibility at both node and workload granularity. Kepler~\cite{10254956,10643925} exports process- and container-level energy metrics in Kubernetes clusters, but it relies on power interfaces such as RAPL, ACPI, or \texttt{hwmon}, which are often unavailable on ARM edge boards such as Raspberry Pis. 
The PowerAPI ecosystem~\cite{fieni:hal-04601379} uses performance counters and calibration data to build container-level software-defined models (e.g., SmartWatts~\cite{9139675}). 
Kasioulis et al.~\cite{10.1007/978-3-031-50684-0_20} model Raspberry Pi power from containerized CPU and network stress workloads and evaluate on two ML inference workloads.
GreenPipe differs by combining external-meter labeling, DNN-oriented multi-resource training, Kubernetes online deployment, and heuristic container attribution for ARM edge nodes.

\paragraph{DNN inference: energy and performance.}
DNN inference is increasingly deployed across edge--cloud systems, where model architecture, precision, parallelism, and serving mode affect both latency and energy~\cite{9596610}. MLPerf Power~\cite{Tschand2024MLPerfPB} standardizes the methodology for measuring inference performance and power, but does not provide deployable per-node power models for Kubernetes edge nodes. 
GreenPipe uses DNN-relevant training benchmarks, validates models on realistic inference workloads, and reports latency--energy behavior for different deployments.

\section{System Overview}

\begin{figure}[hbt]
  \centering
  \begin{minipage}{0.48\linewidth}
    \centering
    \includegraphics[width=\linewidth]{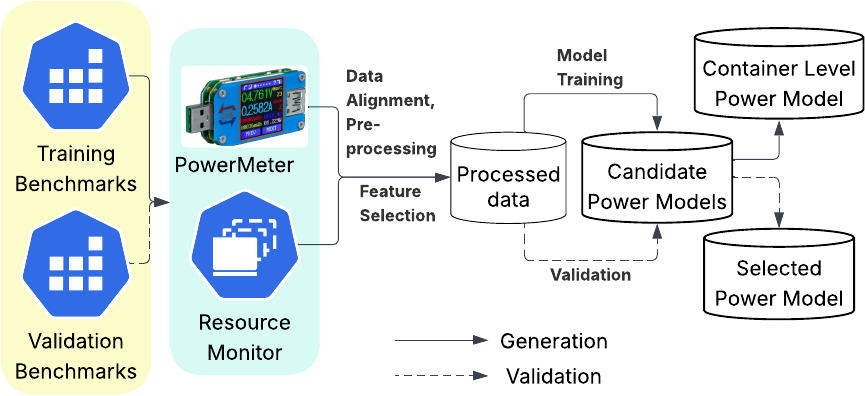}\\[2pt]
    \caption{GreenPipe offline workflow.}
    \label{fig:overview:pipeline}
  \end{minipage}\hfill
  \begin{minipage}{0.48\linewidth}
    \centering
    \includegraphics[width=\linewidth]{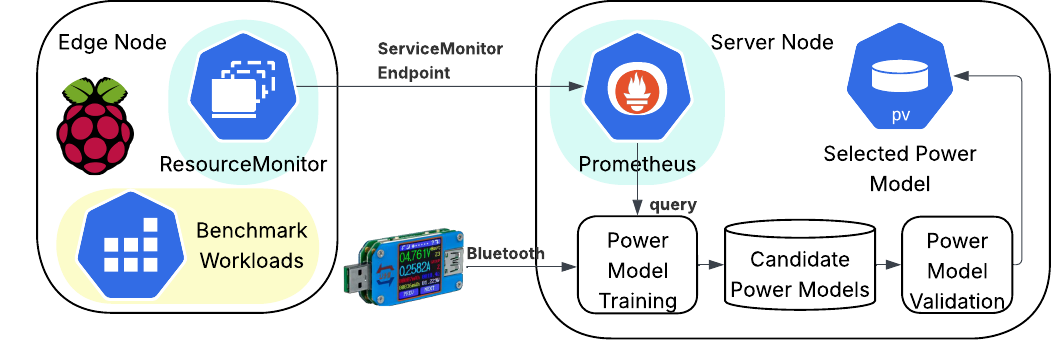}\\[2pt]
    \caption{Deployment of offline workflow: benchmarks running on the edge; data preprocessing, model training, validation on the server.}
    \label{fig:overview:offline-deployment}
  \end{minipage}
\end{figure}

\begin{figure}[htbp]
    \centering
    \includegraphics[width=0.5\linewidth]{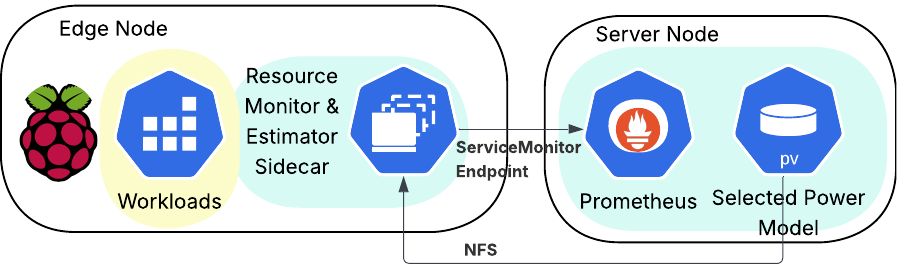}
    \caption{Power model prediction deployment}
    \label{fig:overview:online-deployment}
\end{figure}

GreenPipe operates in two phases: an \textbf{offline workflow} as shown in Figure \ref{fig:overview:pipeline}-\ref{fig:overview:offline-deployment} for model generation and validation, and an \textbf{online operation} phase as shown in Figure \ref{fig:overview:online-deployment} for runtime power estimation.
This section summarizes the main components and their interactions.

\paragraph{Offline Workflow.}
GreenPipe first collects resource and power traces on the edge during containerized training benchmarks (Section~\ref{sec:method:trainingbench}) with an external power meter attached, then pre-processes data, trains and cross-validates regression models on the server. 
The selected candidates are evaluated on validation benchmarks (Section~\ref{sec:setup:validationworkloads})---DNN inference workloads under local and serving modes---before one model is chosen for online deployment. 
Three main components are active during offline workflow: benchmark workloads are launched as Kubernetes Jobs on the edge node; resource monitor exports node-, process-, and container-level metrics and run as DaemonSet; Prometheus is deployed as a StatefulSet on the server node and stores time-series metrics.

\paragraph{Online operation.}\label{sec:overview:online}
After validation, the selected regression power model is stored in a cluster-accessible volume. In addition to the components active during offline workflow, an estimator sidecar that loads the trained model and predicts node-level power at runtime with per-second granularity from live metrics streamed by the monitor is added to the Daemonset. Container-level values are then derived on the edge by attributing predicted node power as described in Section~\ref{sec:method:container}. Predictions are exported to Prometheus for analysis and future energy-aware scheduling.

GreenPipe produces a per-node power model for a given edge hardware configuration. The server acts as orchestrator, metrics backend, and---for served inference---request generator. We focus on reproducible per-node power modeling and performance--energy characterization.

\section{GreenPipe Methodology}

This section introduces the methodology for generating power models for edge nodes in a Kubernetes-based cluster. It begins with the resource and power monitoring methods, training benchmark design, model training and selection procedure and the container attribution approach.

\subsection{Resource and Power Monitoring}

To enable real-time power prediction, our resource collecting system is developed based on Kepler~\cite{10254956}, an open-source Kubernetes exporter that collects power-level metrics. However, Kepler is built for x86 architectures and collects power data based on RAPL, which is not applicable to Raspberry Pis. We customize the resource monitor and rebuild the image on ARM architecture and use an external meter at 1Hz to collect power.

The resource monitor module is written in \textit{Go}, providing different profiling granularity---process, container and node---and collecting different types of metrics related to CPU activity, cache behavior, memory/page-cache activity, disk I/O, and network traffic, which are the main resource dimensions exercised by local and served DNN inference. The components and their sources are summarized in Table~\ref{tab:method:4.1:resources}.
Container-level metrics are aggregated through cgroup identifiers, PIDs and network namespaces when applicable.

During the online phase (Section~\ref{sec:overview:online}) the collected resource usage metrics are passed to the estimator and the per-second power estimation is exported.

\begin{table}[tbh]
    \centering
    \caption{Monitored resources and their obtained sources}
    \begin{tabular}{c|c|c}\hline
    \textbf{Components} & \textbf{Metrics} & \textbf{Source} \\ \hline
       \multirow{2}{*}{CPU}  &  \texttt{cpu\_cycles, cpu\_instructions} & Performance Monitoring Unit (PMU) \\
         & \texttt{cpu\_time}  & eBPF trace via kernel hooks\\ \hline
      \multirow{2}{*}{memory}   & \texttt{cache\_miss} & Performance Monitoring Unit (PMU) \\ 
        & \texttt{page\_cache\_hit} & eBPF trace via kernel hooks \\ \hline
       \multirow{2}{*}{network}   & \texttt{net\_tx\_irq, net\_rx\_irq} & eBPF trace via kernel hooks\\
       & \texttt{network\_bandwidth} & /proc/net/dev \\ \hline
        \multirow{2}{*}{disk} &\texttt{block\_io\_irq} & eBPF trace via kernel hooks\\
        & \texttt{disk\_io} & cgroupPath/io.stat, /proc/diskstats \\
        \hline
    \end{tabular}
    \label{tab:method:4.1:resources}
\end{table}

\subsection{Training Benchmarks}\label{sec:method:trainingbench}
GreenPipe uses three benchmark groups to build training data: micro-benchmarks, combined benchmarks, and DNN inference-targeted benchmarks, summarized in Table \ref{tab:method:4.2:trainingbench}.
All benchmarks are containerized and executed on the edge node.

\begin{table}[htb]
\centering
\caption{Training benchmarks}
\begin{tabular}{l|l|l}
    \hline
\textbf{Benchmark} & \textbf{Workload} & \textbf{Description} \\ 
    \hline
\multirow{6}{*}{stress-ng (v0.18.07)}  & cpu\_int8 & \\ \cline{2-2}
                   & cpu\_fp16 & stress CPU by intensive arithmetic\\ \cline{2-2}
                   & cpu\_fp32 & \\ \cline{2-3}
                   & cpu\_matrixprod & stress CPU by matrix multiplication \\ \cline{2-3}
& vm-rw &  stress memory by read/write operations \\ \cline{2-3}
& hdd & stress disk by read/write operations \\ \hline
iperf (v3.12) & network & network packets transmission/receive \\ \hline
\multirow{2}{*}{Combined} & cpu \& mem \& disk & stress multiple components by stress-ng \\ \cline{2-3}
 & all components & stress all components by stress-ng and iperf \\ \hline
\multirow{3}{*}{DeepBench} & gemm\_bench & mixed precision dense GEMM \\ \cline{2-3}
                   & sparse\_bench & mixed precision sparse GEMM \\ \cline{2-3}
                   & conv\_bench & fp32 precision convolution computation  \\ \hline
\end{tabular}
\label{tab:method:4.2:trainingbench}
\end{table}

\paragraph{Micro-benchmarks} leverage \texttt{stress-ng} and \texttt{iperf} to create loads for different components. 
The results of micro-benchmarks show strong positive correlation of CPU-related features with power consumption. 

\paragraph{Combined benchmarks} are included to reflect practical DNN inference behavior. 
For instance, the workload exercising CPU, memory and disk creates scenarios for offline local inference where loading models requires memory and disk read while storing results requires write.
In addition, workloads that stress CPU, memory, disk and network simultaneously are designed to approximate serving inference scenarios, where the inference requests arrive over the network. 

\paragraph{DNN inference-targeted benchmarks} -- DeepBench\footnote{\url{https://github.com/baidu-research/DeepBench}} contains basic DNN kernel computations at low level, e.g., dense/sparse matrix multiplies and convolution computations. These workloads do not rely on deep learning frameworks or models built for applications to capture the DNN inference behavior. For each benchmark, it computes for different sizes of matrices and convolutions at least 10 times and 50 times, respectively.

\subsection{Data Preparation, Model Training and Selection}

After running the training benchmarks, the collected resource metrics and meter power data are aligned by timestamps at 1 Hz. Samples with missing metric or power values are removed, and features are normalized before model fitting. 
Based on correlation analysis, seven representative features are selected: \texttt{cpu\_cycles, cpu\_time, cpu\_instructions, cache\_miss, page\_cache\_hit, net\_bandwidth, disk\_io}. 
At the end, the training dataset is obtained with 13938 samples generated by 12 benchmarks.

We treat power estimation as a supervised regression problem. GreenPipe compares representative linear and non-linear regressors from scikit-learn, including Linear Regression (LR), Lasso Regression (Lasso), Ridge Regression (Ridge), Linear Regression fitted with Stochastic Gradient Descent (SGD), Polynomial Regression (PR), Decision Tree (DT), Random Forest (RF), Gradient Boosting (GB), eXtreme Gradient Boosting (XGB), K-Nearest Neighbors (KNN), and Support Vector Regression (SVR). 
Our goal is not to introduce a new regressor, but to select a robust model under the same training data. 
Models are selected according to validation MAPE; the validation benchmarks are introduced in Section~\ref{sec:setup:validationworkloads}.

\subsection{Container-level Power Model}\label{sec:method:container}

We heuristically attribute node-level power to containers by decomposing node power into idle and dynamic components. The node idle power
$P_{\mathrm{node}}^{\mathrm{idle}}$ is estimated as the steady-state minimum power observed during idle periods, and the remaining part is treated as dynamic power:
\begin{equation}
P_{\mathrm{node}}^{\mathrm{dyn}} = P_{\mathrm{node}} - P_{\mathrm{node}}^{\mathrm{idle}}
\end{equation}
This follows the common static/dynamic power decomposition used in software power meters~\cite{9139675,10643925}.
For each container, dynamic power is attributed proportionally to its resource usage:
\begin{equation}
P_{\mathrm{container}}^{\mathrm{dyn}} =
\frac{R_{\mathrm{container}}}{R_{\mathrm{node}}} \cdot
P_{\mathrm{node}}^{\mathrm{dyn}}
\end{equation}
Here, $R$ denotes a weighted sum of selected resource metrics,
$R = \sum_i w_i R_i$, where the weights $w_i$ are obtained from the trained LR model. We aggregate resources ensuring
$\sum_c R_c = R_{\mathrm{node}}$ over the containers on the node. 
Idle power is attributed according to the number of processes in each container:
\begin{equation}
P_{\mathrm{container}}^{\mathrm{idle}} =
\frac{N_{\mathrm{proc,container}}}{N_{\mathrm{proc,node}}} \cdot
P_{\mathrm{node}}^{\mathrm{idle}}
\end{equation}
The final container estimate is
$P_{\mathrm{container}} =
P_{\mathrm{container}}^{\mathrm{dyn}} +
P_{\mathrm{container}}^{\mathrm{idle}}$. 

This attribution is a heuristic for operational visibility. Since the external meter provides only node-level ground truth, we do not claim validated per-container power accuracy.

\section{Experimental setup}

\subsection{Experimental Settings}

\paragraph{Hardware.} 
The server node is equipped with 8 x Intel Core i7-8650U CPU @ 1.90GHz. 
The edge node is a Raspberry Pi 4 Model B Rev 1.5 equipped with Broadcom BCM2711, Quad core Cortex-A72 (ARM v8) 64-bit SoC @ 1.8GHz, and an integrated dual-band (2.4 GHz/5 GHz) IEEE 802.11ac Wi-Fi interface. 
The power meter is the Ruideng UM25C USB meter.

\paragraph{Platform Settings.} 
The server node runs Ubuntu 22.04 with 6.8.0-85-generic kernel. 
The edge node runs Debian GNU/Linux 12 with 6.6.56-v8+ kernel. 
We customized and rebuilt the kernel to enable eBPF attachment to specific events on Raspberry Pi.  
The edge-server cluster is deployed using K3s\footnote{\url{https://k3s.io/}} v1.30.3+k3s1, a lightweight Kubernetes distribution designed for resource-constrained environments, in which the server node runs the k3s server and the edge node runs the k3s agent. 
The container runtime used in the experiments is \textit{containerd}://1.7.17-k3s1; cgroup version is v2 to enforce resource isolation for containerized workloads. 
Prometheus (v2.53.1) is deployed with prometheus-operator (v0.75.2) to scrape metrics profiled by service monitor and store them as time-series. 
The DVFS governor on the edge node is \textit{ondemand}.

\subsection{Validation benchmarks}\label{sec:setup:validationworkloads}

We use Image Classification as the validation task considering the practical workloads deployed at the edge. 
Vision models for this inference task include MobileNetV2, EfficientNetB0, and ResNetV2.
These models and pretrained weights are obtained from Keras\footnote{\url{https://keras.io/api/applications/}} and transformed into \texttt{saved\_model.pb} with float32 (noted as \texttt{pb} model).
Additionally, we generated light models (noted as \texttt{tflite} models) with float32 precision and quantized versions (int8 and float16) by LiteRT converter to evaluate the performance of different types of CPU calculations.
To conduct the quantization, we use the calibration dataset\footnote{\url{https://github.com/mlcommons/inference/blob/master/calibration/ImageNet/cal\_image\_list\_option\_2.txt}}
 provided by MLPerf. 
The dataset for the inference task is sampled from ImageNet (Large Scale Visual Recognition Challenge, ILSVRC2012) validation set.

Our on-device inference scenarios consist of \textit{local} inference and \textit{serving} inference. 
Local scenario represents applications inferring local data that are already collected on device and usually processed in batches. 
This scenario is implemented by LiteRT inference engine for \texttt{tflite} models. 
Serving scenario represents online inference services deployed on the edge node where data are sent from other nodes on the fly. Requests are sent to the endpoint through gRPC API. This is implemented by TensorFlow Serving. 

Note that no validation inference trace is used for model training. Reported errors are computed over all aligned samples in the validation traces. 

\section{Evaluation}

In this section, we evaluate the power models on different DNN inference workloads and deployment scenarios.
The metrics for evaluating prediction accuracy are mean absolute error ($MAE=\frac{1}{n}\sum^n_{i=1}|y_i-\hat{y}|$, unit in Watt) and mean absolute percentage error ($MAPE=\frac{1}{n}\sum^n_{i=1}|\frac{y_i-\hat{y}}{y_i}|*100\%$).
We first compare our GreenPipe generated power models with the baseline power models, then analyse the impact of deployment and workload factors on power model prediction. We additionally provide the measured latency--energy trade-offs and show the results of online deployment and container power attribution.

\subsection{System-level power model accuracy and comparison}

We compare the power predicted by our GreenPipe with Baseline Training~\cite{10994218} that collects the training dataset by only stressing CPU-related components, and Baseline LR~\cite{KANSO2023100837} $P=4.5344\times U+2.2857$ using only CPU utilization calculated by \texttt{cpu\_time} on our validation workloads. The prediction results of all the candidate power models are summarized in Table \ref{tab:eval:6.1:overall}.

$MAPE$ of GreenPipe-trained power models ranges from 6.3\% to 9.4\%. 
Compared with Baseline Training, GreenPipe reduces $MAE$ by 46.9\% and $MAPE$ by 26.9\% on average, showing the validity and efficacy of comprehensive training benchmarks.
All the GreenPipe power models improve over the Baseline LR, indicating the considerable importance of non-CPU centric features for power modeling. 
Across those regressors, differences in inference validation are modest compared with the gap to CPU-stress and utilization-only baselines, confirming that multi-resource benchmark coverage matters more than the choice of regressor.

\begin{table}[tb]
    \centering
    \caption{Power model prediction results on validation benchmarks}
    \begin{tabular}{l|l|c|c|c|c|c|c|c|c|c|c|c|c}\hline
    \multicolumn{2}{c|}{Regressor} & LR & Ridge & Lasso & SGD &PR &DT&RF&GB&XGB & SVR & KNN & BaselineLR\\ \hline
    Baseline & $MAE$ & 0.59 &0.58 & 0.60 & 0.60 & 0.72 & 0.54 & 0.59 & 0.60 & 0.58 & 0.76 & 0.55 & 0.71 \\ \cline{2-14}
    Training & $MAPE$ & 10.8 & 13.1 & 13.9 & 12.8 & 13.3 & 8.8 & 10.4 & 11.3 & 11.1 & 13.2 & 9.8 & 14.2 \\ \hline
    \multirow{2}{*}{GreenPipe } & $MAE$ & 0.36 & 0.37 & 0.39 & \textbf{0.22} & 0.29 & 0.24 & 0.33 & 0.40 & 0.39 & 0.49 & 0.23 & - \\ \cline{2-14}
         & $MAPE$ & 8.2 & 8.4 & 8.9 & 8.5 & \textbf{6.3} & 7.3 & 8.4 & 6.9 & 8.9 & 9.4 & 7.0 & - \\ \hline 
    \end{tabular}
    \label{tab:eval:6.1:overall}
\end{table}

\subsection{Impact of deployment and workload factors}\label{sec:eval:factors}
Our validation workloads are categorized into several groups: workloads with different DNN architectures and precisions, number of threads doing inference, inference engines and scenarios.
In this section, we study the impact of these workload factors on power model accuracy within these groups.

\paragraph{DNN model architecture and precision.}

Among the three distinct DNN models, the power predictions of MobileNet inference workloads are the least accurate ($MAPE$ from 6.6\% to 9.2\%) compared to the other bigger models.
Our power models predict better in floating-point inference than integer inference.
Besides, non-linear models have smaller errors in prediction than the linear models, indicating the non-linear relation between selected features and power. 

\paragraph{Thread counts.}

\begin{figure}[tbhp]
  \centering
  \begin{minipage}{0.48\linewidth}
    \centering
    \includegraphics[width=\linewidth]{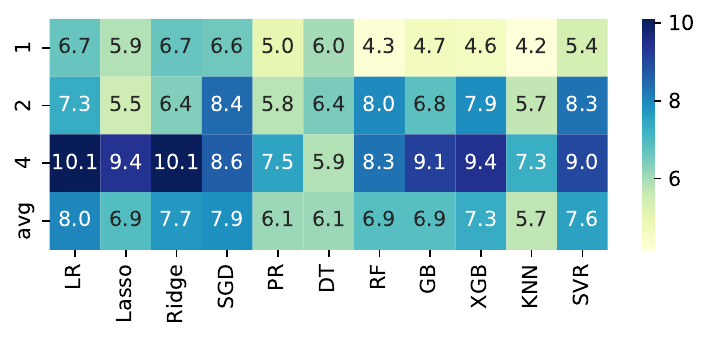}\\[2pt]
    \caption{MAPE of power prediction on inference workloads with different threads.}
    \label{fig:eval:thread}
  \end{minipage}
  \hfill
  \begin{minipage}{0.48\linewidth}
    \centering
    \includegraphics[width=\linewidth]{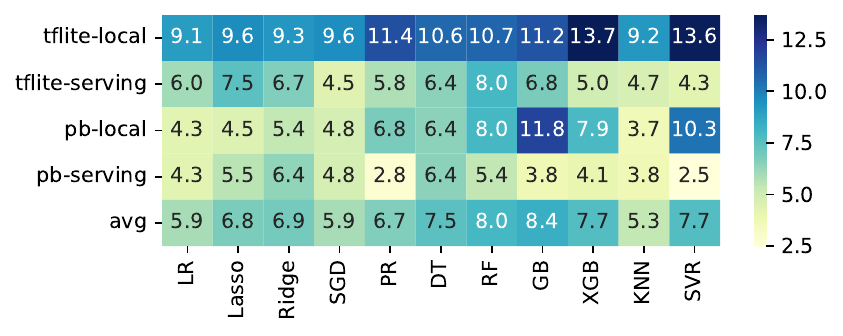}\\[2pt]
    \caption{MAPE of power prediction on inference workloads with different inference engines and scenarios.}
    \label{fig:eval:scenario}
  \end{minipage}
\end{figure}

As shown in Fig. \ref{fig:eval:thread}, for inference workloads conducted with different numbers of threads, power prediction accuracy decreases as the number of threads increases.
Both the linear and non-linear models provide reasonable power predictions.

\paragraph{Inference engine/scenarios.}

Fig. \ref{fig:eval:scenario} shows the comparison of the power prediction for different inference engines and inference scenarios. It is significant that the power model results for \texttt{pb} workloads are better than those for \texttt{tflite} workloads. The actual power of \texttt{tflite} workloads are consistently slightly larger than the predicted values. This may be because LiteRT inference engine is more optimized for edge devices than TensorFlow.

These results show that GreenPipe remains effective across deployment schemes. Among workload factors, precision and model architecture have a limited impact on prediction error, whereas numbers of threads, inference engines and deployment modes (local vs. network-served) dominate prediction difficulty.

\subsection{DNN workloads latency--energy trade-offs}

We characterize performance--energy trade-offs for validation workloads using measured latency and ground-truth node energy computed by integrating meter power during the inference, $E=\int P_{gt}\,dt$.
Among the models, ResNet exhibits the highest latency and energy. Among the precisions, int8 is the fastest and less energy consuming. 
The effect of thread count is model-dependent. Smaller models benefit more from parallelism in both latency and energy; while for larger models, it mainly reduces latency and the higher power draw from using more threads offset the shorter runtime and lead to higher energy consumption.
Overall, precision and model family induce smaller shifts in the latency--energy frontier than deployment mode and thread count (cf.\ Section~\ref{sec:eval:factors}), suggesting that energy-aware placement should prioritize parallelism before model metadata alone.
These results complement power prediction accuracy by quantifying physical performance--energy behavior on the edge node.

\begin{figure}[tbhp]
  \centering
  \begin{minipage}{0.48\linewidth}
    \centering
    \includegraphics[width=\linewidth]{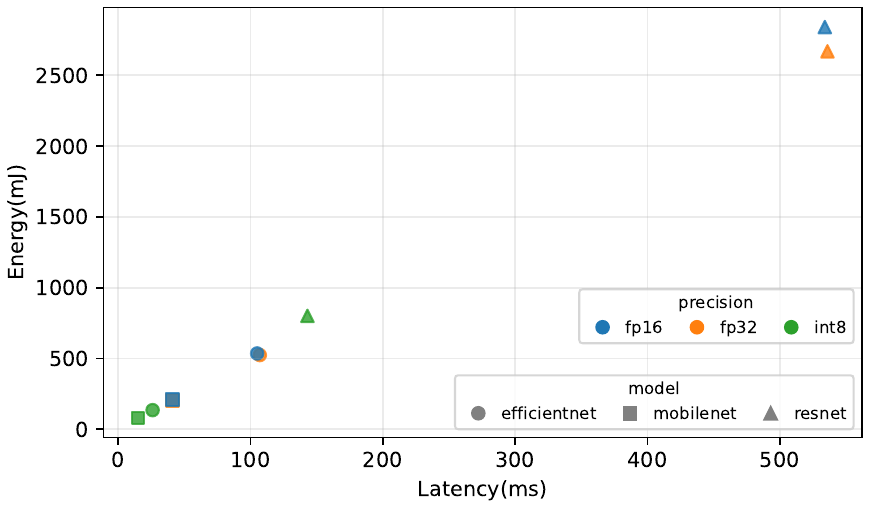}\\[2pt]
    \caption{Energy-latency trade-off between models and precisions. Scenario: \texttt{tflite} running locally with thread 1.}
    \label{fig:tradeoff:precision}
  \end{minipage}
  \hfill
  \begin{minipage}{0.48\linewidth}
    \centering
    \includegraphics[width=\linewidth]{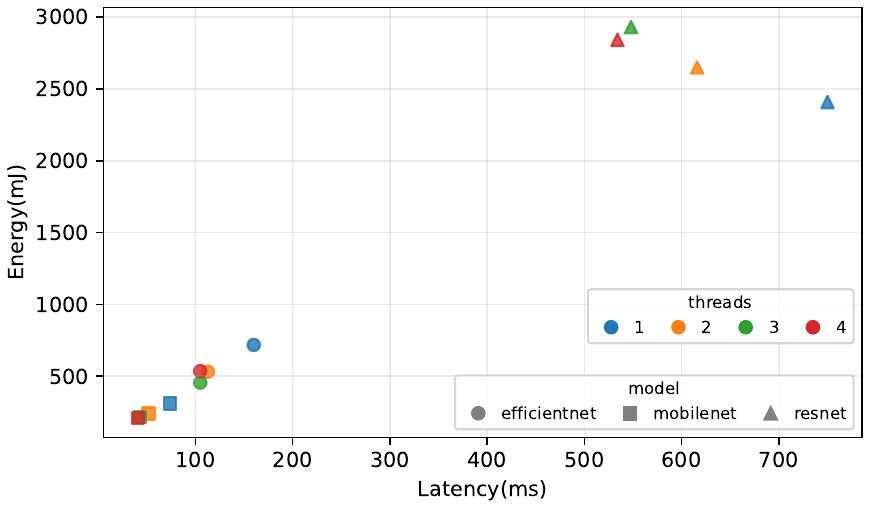}\\[2pt]
    \caption{Energy-latency trade-off between models and threads. Scenario: \texttt{tflite} running locally with fp16 precision.}
    \label{fig:tradeoff:thread}
  \end{minipage}
\end{figure}

\subsection{Online power estimation and container-level attribution}
\begin{figure}[ht]
        \centering
    \includegraphics[width=0.75\linewidth]{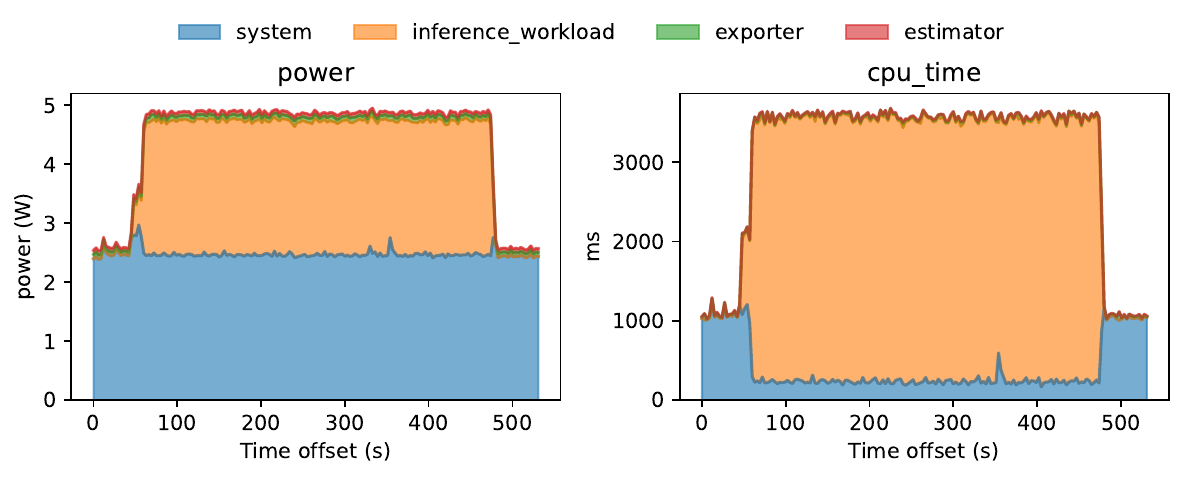}
    \caption{Container-level power and \texttt{cpu\_time} attribution during online operation, including the inference workload, exporter, estimator containers and remaining system activity. Scenario: MobileNetV2 \texttt{pb} running locally with 4 threads.}
    \label{fig:eval:container}
\end{figure}

This experiment illustrates runtime attribution rather than validating container-level accuracy, because the external meter only provides node-level ground truth. 
Fig. \ref{fig:eval:container} presents the container-level attributed power (left) and one of the monitored resources, \texttt{cpu\_time} (right), of an example inference workload. The node-level power is predicted by the trained LR model and then attributed to containers using the method in Section~\ref{sec:method:container}.
In this scenario, most \texttt{cpu\_time} is consumed by the running workloads. Before the workload starts, the system remains idle; after the computation-intensive workload begins, the inference container dominates CPU usage. 
The power distribution differs from \texttt{cpu\_time} because the power model accounts for idle power, dynamic power and multiple resource metrics. 
Background system power, including standby activity, kernel execution, clocks, and DRAM refresh, is treated as an approximately constant component by the model. 
The exporter and estimator containers together account for less than 2\% of total node power, indicating that the online monitoring and estimation overhead is small in this experiment.

\section{Conclusion and Future Work}

In this work, we presented GreenPipe, a data-driven pipeline for building and deploying per-node power models for containerized DNN inference on ARM edge devices. 
GreenPipe integrates multi-resource profiling, benchmark-driven model training, and online estimation in a Kubernetes cluster, with heuristic container-level attribution derived from node-level predictions. 
On a Raspberry Pi 4 edge node, GreenPipe achieves system-level prediction errors of 6.3\%--9.4\% $MAPE$ on validation inference workloads. We also report latency--energy trade-offs across workload configurations, showing that thread-level parallelism can reduce latency without always reducing energy. 
Future work will extend the study to additional edge platforms, GPU-equipped devices, and broader DVFS settings. Since online deployment does not provide synchronous ground-truth power labels, we will also investigate periodic meter-assisted recalibration when new labeled data are available.

\subsubsection{\ackname}
This work was partially supported by Spanish Ministry of Science (MICINN), the Research State Agency (AEI) and European Regional Development Funds (ERDF/FEDER) under contract PID2024-160996OB-I00, MICIU/AEI/10.13039/501100011033, and by the Generalitat de Catalunya (AGAUR) under contract 2021-SGR-00478.
This work was also supported in part by the Norwegian Research
Council under Grant 322473 (AirQMan project), as well as by the EU’s Digital Europe Programme under Grant 101123471 (EDGE-Skills project).

\bibliographystyle{IEEEtran}
\bibliography{bib}

\end{document}